\documentclass[aps,showpacs,12pt]{revtex4}
\usepackage{dcolumn}
\usepackage{graphicx}
\usepackage{amsmath}
\usepackage{amsfonts}
\usepackage{amssymb}
\usepackage{psfrag}
\usepackage{wrapfig}
\usepackage{subfigure}
\usepackage{makeidx}
\usepackage{bm}
\usepackage{epsf}
\usepackage{hyperref}
\usepackage{color}
\usepackage{multirow}
\usepackage{mathrsfs}
\usepackage{mathrsfs}

\begin{document}

\title{Gravito-Electromagnetic Perturbations and QNMs of Regular Black Holes}

\author{Kun Meng$^1$}
\email{kunmeng@wfu.edu.cn}
\author{Shao-Jun Zhang$^{2,3}$}
\email{sjzhang@zjut.edu.cn}
\affiliation{$^1$School of Physics and Electronic Information, Weifang University, Weifang 261061, China\\
 $^2$Institute for Theoretical Physics $\&$ Cosmology, Zhejiang University of Technology, Hangzhou 310032, China\\
	$^3$United Center for Gravitational Wave Physics, Zhejiang University of Technology, Hangzhou 310032, China}
	
\date{\today}                             

\begin{abstract}
In the framework of Einstein's gravity coupled to nonlinear electromagnetic fields, we study gravito-electromagnetic perturbations of magnetic regular black holes. The master equations of perturbations are obtained through Chandrasekhar's procedure, in which gravitational perturbations with odd-parity are coupled to the electromagnetic perturbations with even-parity. As an application, we apply the master equations to obtain quasinormal modes (QNMs) for three types of regular black holes by using numerical method. Results show that QNMs of regular black holes depends significantly on the parameters of the theory and the magnetic charge of the black holes and are very different from that of the Reissner-Nordstr\"om black hole. Indications of these results on the stability of these regular black holes are discussed in detail.
\end{abstract}

\maketitle

\section{Introduction}
In the past few years, significant astronomical discoveries on black holes have been made, including the first-ever observations of gravitational waves (GWs) from coalescence of binary black holes\cite{1602.03837,1606.04855} and the first images of the M87* and SgrA* shadows\cite{1906.11238,EventHorizonTelescope:2022xnr}, confirming the existence of black holes (BHs) in the Universe overwhelmingly and opening the new era of astronomical detection. Because of the long range nature of gravity, GW channel has incomparable advatanges in astronomical observations providing us a noval detection method beyond the traditional optical method. With the
continuous improvement of detection ability of GWs, we are now able to test general relativity (GR) in the strong gravity regime with unprecedented precision by examining phenomena around BHs.

According to GR, BH will be the inevitable end state of the collapse of very massive stars \cite{Tolman,Oppennheimer,Joshi,Christodoulou,Joshi93,Penrose}. Penrose and Hawking proved that, under the strong energy condition, singularity will appear inevitably during the collapse \cite{Penrose:1964wq,HawkingEllis}. However, it is widely believed that singularity is not physical and should
be resolved by, for example, taking into account the quantum effects of either matter or gravitational field. For the electromagnetic field, one effective way to include its quantum effects is to consider its nonlinear extensions. Actually, there has been a long history on the study of nonlinear electromagnetic field theories. Early in 1930s', Heisenberg and Euler showed that higher order corrections to Maxwell theory arose when considering vacuum polarization\cite{0605038}, and Born and Infeld constructed an electromagnetic field theory tempting to obtain a finite self-energy of the electron\cite{BI}. Subsequent researches found that Born-Infeld (BI) electromagnetic theory also arises in the low energy effective theory of string theory \cite{Polchinski:1998rq,Polchinski:1998rr}. Various BH solutions of gravity coupled to BI electromagnetic fields were found and studied extensively in past years \cite{Hoffmann,0306120,0406169,0410158,0802.2637,1712.08798,1804.10951,2102.05112,2201.09703}. More interestingly, regular BH solutions with no essential singularity are indeed found to exist in this framework. Bardeen first attempted to construct BHs free of singularity \cite{Bardeen}. Later, Ay\'o{}n–Beato and Garc\'\i{}a successfully constructed exact regular BH solutions of Einstein equations\cite{9911046,9911174}. More recently, Fan and Wang constructed a new class of general regular BH solutions \cite{1610.02636}. For more interesting work on this topic, see also Refs. \cite{Cai:2020kue,Cai:2021ele}. Then it is nature to ask the following questions: Do regular BHs indeed exist or they are just mathematical outcomes? From the aspect of astrophysical observations, is there any possible signal to distinguish regular BHs from the standard BHs?

It is now well known that, in the late-time (the so-called ringdown stage) of the coalescence of binary BHs (or BH-neutron star, binary neutron stars), GW signal emitted is a superposition of series of sinusoidal damping modes. These modes, called QNMs, have characteristic frequencies determined uniquely by parameters of the final formed BH, namely its mass, angular momentum and electromagnetic charge (if present). Also, QNMs depends on the specific theory underlying the BH. Thus it provides us a powerful tool to distinguish different BHs and theories by detecting QNMs in GWs. Amounts of effort have been devoted on studying QNMs of various BHs in variety of theories in past years, see reviews \cite{Kokkotas:1999bd,Nollert:1999ji,Berti:2009kk,Konoplya:2011qq}. Although limited by detection accurary, data analysis to read off QNMs from the detected GW signals are ongoing and some progresses have been made \cite{LIGOScientific:2020tif,Isi:2019aib,Isi:2020tac,Capano:2021etf,Bustillo:2020buq,Cotesta:2022pci,Isi:2022mhy,Finch:2022ynt}. It is expected that the next generation of detectors can give us more
conclusive information about QNMs of BHs. Study of QNMs of regular BHs have also attracted lots of attention \cite{Cai:2020kue,Cai:2021ele,Saleh:2018hba,1805.00240,1807.09065,1810.06383,0706.1513,1208.5442,1703.04286,1811.02847,Lin:2013ofa,2108.07766,2109.11878,2111.06488,Bronnikov:2012ch,Konoplya:2022hll}, and results show that regular structure of BH origin may have considerable influences on QNMs. However, previous studies just consider either gravitational or matter perturbations or just consider test fields, investigations of gravitational and matter perturbations simultaneously in a consistent way are rare \cite{0208090,2004.07560,Daghigh:2021psm,2004.12185,2111.06273} and are urgently needed. So, in this paper, our aim is to study gravitational and electromagnetic perturbations of regular BHs simultaneously. By applying Chandrasekhar’s procedure\cite{Chandrasekhar}, general master perturbation equations are obtained which are coupled Sch\"odinger-like equations. As applications of the master equations, we then calculate fundamental QNMs of three types of regular BHs constructed in \cite{1610.02636} by numerical method. We hope that this study can help to answer the questions mentioned above.

This paper is organized as follows. In Sec. \ref{section2}, we obtain the first-order master perturbation equations, and rewrite them into the standard form through separating variables and introducing proper new functions. In Sec. \ref{section3}, we apply the master perturbation equations to obtain fundamental QNMs for three types of regular BHs by numerical method, and discuss their physical indications. The last section is devoted to summary and discussions.

\section{Gravito-electromagnetic perturbations\label{section2}}
In this section, we perturb the field equations of Einstein gravity coupled to nonlinear electromagnetic field theory, and to obtain first-order perturbation equations. Through separating variables and defining new functions we successfully rewrite the perturbation equations into the standard Sch\"odinger-type form at last.
\subsection{Field equations}
The action which describes Einstein gravity coupled to nonlinear electromagnetic field reads\cite{1610.02636}
\begin{align}
I=\frac{1}{16\pi}\int \mathrm{d}^4x\sqrt{-g}\left(R-\mathcal{L}(\mathcal{F})\right),
\end{align}
with
\begin{align}
\mathcal{L}(\mathcal{F})=\frac{4\mu}{\alpha}\frac{(\alpha \mathcal{F})^{\frac{\nu+3}{4}}}{\left(1+(\alpha \mathcal{F})^{\frac{\nu}{4}}\right)
^{\frac{\mu+\nu}{\nu}}},
\end{align}
where $\mathcal{F}\equiv F_{\mu\nu}F^{\mu\nu}$ and $F_{\mu\nu}$ is the field strength tensor. Standard variation process leads to the equations of motion (EOMs) for the metric
\begin{align}
R_{\mu\nu}-\frac{1}{2}Rg_{\mu\nu}=2\left(\mathcal{L}_{\mathcal{F}}F^2_{\mu\nu}-\frac{1}{4}g_{\mu\nu}\mathcal{L}\right),
\end{align}
and the electromagnetic fields
\begin{align}
\nabla_\mu\left(\mathcal{L}_{\mathcal{F}}F^{\mu\nu}\right)=0\label{eomEM1},
\end{align}
respectively. A general class of magnetic regular BH solutions were found in Ref.\cite{1610.02636}
\begin{align}
&\mathrm{d}s^2=-e^{2\gamma}\mathrm{d}t^2+e^{-2\gamma}\mathrm{d}r^2+r^2\mathrm{d}\Omega^2,\;\;\;\;\;A=Q_m\cos\theta \mathrm{d}\varphi,\nonumber\\
&e^{2\gamma}=1-\frac{2Mr^{\mu-1}}{(r^\nu+q^\nu)^{\mu/\nu}},\label{magbh}
\end{align}
where $q$ is the magnetic charge and $\mu, \nu$ are free parameters. Through calculating curvature invariants, it's concluded that the BHs are free of singularity when $\mu\geq3$. Thermodynamic quantities of the regular BHs are given by
\begin{align}
&M_{ADM}=M,\;\;S=\pi r_0^2,\;\; Q_m=\frac{q^2}{\sqrt{2\alpha}},\nonumber\\
&T=\frac{1}{4\pi r_0}\left(1-2\mu\alpha^{-1}q^4r_0^{\mu-1}(r_0+q)^{-\mu-1}\right),\nonumber\\
&\Psi=-\frac{q}{\sqrt{2\alpha}}\left((3-(\mu-3)\frac{q}{r_0})(1+\frac{q}{r_0})^{-\mu-1}-3\right),\nonumber\\
&\Pi=\frac{q^3}{4\alpha^2}\left((1+(\mu+1)\frac{q}{r_0}(1+\frac{q}{r_0})^{-\mu-1}-1\right).
\end{align}
It is straightforward to check that the first law of thermodynamics
\begin{align}
dM_{ADM}=TdS+\Psi dQ_m+\Pi d\alpha
\end{align}
and the Smarr formula
\begin{align}
M_{ADM}=2TS+\Psi Q_m+2\Pi  \alpha,
\end{align}
are satisfied.

To study gravito-electromagnetic perturbations of the background black hole spacetime, we adopt Chandrasekhar's procedure\cite{Chandrasekhar}. First, we choose a metric with sufficient generality to describes a non-stationary and axisymmetric spacetime
\begin{align}
\mathrm{d}s^2=-e^{2\gamma}(\mathrm{d}t)^2+e^{2\psi}(\mathrm{d}\varphi-q_2\mathrm{d}x^2-q_3\mathrm{d}x^3-\sigma\mathrm{d}t)^2+
e^{2\mu_2}(\mathrm{d}x^2)^2+e^{2\mu_3}(\mathrm{d}x^3)^2.\label{metric1}
\end{align}
Note that the metric (\ref{metric1}) involves seven quantities $\gamma, \psi, \mu_2, \mu_3, \omega, q_2, q_3$, which are functions of $t, x^2$ and $x^3$. While Einstein's equation contains only six independent equations, thus among the seven functions there are only six independent ones. It can be shown that the functions are related through
\begin{align}
(\sigma_{,2}-q_{2,0})_{,3}-(\sigma_{,3}-q_{3,0})_{,2}+(q_{2,3}-q_{3,2})_{,0}=0.
\end{align}
The form of metric (\ref{metric1}) can also be used to describe a class of non-stationary and non-axisymmetric spacetime by setting $g^{\mu\nu}(t,\varphi,x^2,x^3)
=g^{\mu\nu}(t,x^2,x^3)h(\varphi)$, where the seven quantities above are functions of all four variables $t, \varphi, x^2, x^3$. Anyway, the metric (\ref{metric1}) is of sufficient generality for the questions we discussed in this paper.

Following Chandrasekhar's procedure, for the gravitational sector of the field equations, we apply Cartan's equations of structure to calculate Riemann and Ricci tensors associated to the metric (\ref{metric1}), while for the matter sector we apply tetrad formalism to work out the relevant contributions to EOMs. In order to apply Cartan's formulation, we take the following basis of one-forms
\begin{align}
\omega^0=e^{\gamma}\mathrm{d}t,\;\;\;\omega^1=e^{\psi}(\mathrm{d}\varphi-q_2\mathrm{d}x^2-q_3\mathrm{d}x^3-\sigma \mathrm{d}t),\;\;\;\omega^2=e^{\mu_2}\mathrm{d}x^2,\;\;\;\omega^3=e^{\mu_3}\mathrm{d}x^3.
\end{align}
In this paper, we use Latin letters for the tetrad indices and Greek letters for the tensor indices. From Cartan's first equation of structure
\begin{align}
\mathrm{d}\omega^i+\omega^i_{\;j}\wedge\omega^j=0,
\end{align}
one is able to calculate the connection one-forms $\omega^i_{\;j}$, and from Cartan's second equation of structure
\begin{align}
\frac{1}{2}R^i_{\;jkl}\omega^k\wedge\omega^l=\Omega^i_{\;\;j}=\mathrm{d}\omega^i_{\;j}+\omega^i_{\;k}\wedge\omega^k_{\;j},
\end{align}
the Riemann tensor can be worked out. The explicit expressions of spin connection and Riemann tensor can be found in Ref.\cite{Chandrasekhar}, so we don't list here. It's straightforward to obtain Ricci tensor from Riemann tensor.

To apply the tetrad formalism, we choose the basis of the orthonormal frame as
\begin{align}
e_{0\mu}&=(-e^\gamma,\;\;\;\;\;\;0,\;\;\;\;\;\;\;\;0,\;\;\;\;\;\;\;\;\;0),\nonumber\\
e_{1\mu}&=(-\sigma e^\psi,\;\;\;e^\psi,\;\;\;-q_2 e^\psi,\;\;\;-q_3 e^\psi),\nonumber\\
e_{2\mu}&=(0,\;\;\;\;\;\;\;\;\;\;0,\;\;\;\;\;\;\;\;\;e^{\mu_2},\;\;\;\;\;\;0),\nonumber\\
e_{3\mu}&=(0,\;\;\;\;\;\;\;\;\;\;0,\;\;\;\;\;\;\;\;\;0,\;\;\;\;\;\;\;\;\;e^{\mu_3}).
\end{align}
The corresponding contravariant vectors are given by
\begin{align}
e_{0}^{\;\mu}&=(e^{-\gamma},\;\;\;\;\;\;\sigma e^{-\gamma},\;\;\;\;\;\;\;0,\;\;\;\;\;\;\;\;\;0),\nonumber\\
e_{1}^{\;\mu}&=(0,\;\;\;\;\;\;\;\;\;e^{-\psi},\;\;\;\;\;\;\;\;\;0,\;\;\;\;\;\;\;\;\;0),\nonumber\\
e_{2}^{\;\mu}&=(0,\;\;\;\;\;\;\;\;\;q_2 e^{-\mu_2},\;\;\;\;\;e^{-\mu_2},\;\;\;\;\;0),\nonumber\\
e_{3}^{\;\mu}&=(0,\;\;\;\;\;\;\;\;\;q_3 e^{-\mu_3},\;\;\;\;\;\;0,\;\;\;\;\;\;e^{-\mu_3}).
\end{align}
It's easy to see that $e_{a\mu}e_{b}^{\;\mu}=\eta_{ab}=diag(-1,1,1,1)$, thus the chosen frame is indeed a locally inertial frame.

For convenience we introduce the tensor
\begin{align}
P_{ab}=e_a^{\;\mu} e_b^{\;\nu} \mathcal{L}_FF_{\mu\nu},
\end{align}
with which EOM of the electromagnetic fields (\ref{eomEM1}) can now be rewritten as
\begin{align}
\eta^{m n}P_{a n|m}=0\label{eommatter1},
\end{align}
where $P_{a n|m}=e_a^{\;\mu} e_n^{\;\nu} e_m^{\;\rho}P_{\mu\nu;\rho}$, and $;$ denotes covariant derivative. It's easy to note that, for the tetrad component of strength tensor the Bianchi identity
\begin{align}
F_{[a b|c]}=0,\label{eommatter2}
\end{align}
is still satisfied. After some lengthy but straightforward calculations, Eqs.(\ref{eommatter1}) and (\ref{eommatter2}) reduce to the following set of eight equations
\begin{align}
&\mathscr{D}_3(e^{\psi+\mu_2}F_{12})-\mathscr{D}_2(e^{\psi+\mu_3}F_{13})+(e^{\mu_2+\mu_3}F_{23})_{,1}=0,\label{eommatter31}\\
&\mathscr{D}_2(e^{\psi+\gamma}F_{01})+\mathscr{D}_0(e^{\psi+\mu_2}F_{12})-(e^{\gamma+\mu_2}F_{02})_{,1}=0,\\
&\mathscr{D}_3(e^{\psi+\gamma}F_{01})+\mathscr{D}_0(e^{\psi+\mu_3}F_{13})-(e^{\gamma+\mu_3}F_{03})_{,1}=0,\\
&(e^{\mu_2+\mu_3}P_{01})_{:0}+(e^{\gamma+\mu_3}P_{12})_{:2}+(e^{\gamma+\mu_2}P_{13})_{:3}\nonumber\\
&\;\;=e^{\psi+\mu_3}P_{02}Q_{02}+e^{\psi+\mu_2}P_{03}Q_{03}-e^{\psi+\gamma}P_{23}Q_{23}\\
&\mathscr{D}_2(e^{\psi+\mu_3}P_{02})+\mathscr{D}_3(e^{\psi+\mu_2}P_{03})+(e^{\mu_2+\mu_3}P_{01})_{,1}=0,\\
&\mathscr{D}_0(e^{\psi+\mu_2}P_{03})-\mathscr{D}_2(e^{\psi+\gamma}P_{23})-(e^{\mu_2+\gamma}P_{13})_{,1}=0,\\
&\mathscr{D}_0(e^{\psi+\mu_3}P_{02})+\mathscr{D}_3(e^{\psi+\gamma}P_{23})-(e^{\mu_3+\gamma}P_{12})_{,1}=0,\\
&(e^{\gamma+\mu_2}F_{02})_{:3}-(e^{\gamma+\mu_3}F_{03})_{:2}+(e^{\mu_2+\mu_3}F_{23})_{:0}\nonumber\\
&\;\;=e^{\psi+\gamma}F_{01}Q_{23}+e^{\psi+\mu_2}F_{12}Q_{03}-e^{\psi+\mu_3}F_{13}Q_{02}\label{eommatter38},
\end{align}
where the notations $f_{:A}\equiv f_{,A}+q_Af_{,1}, Q_{AB}\equiv q_{A:B}-q_{B:A}, \mathscr{D}_A f\equiv f_{:A}+q_{A,1}f=f_{,A}+(q_Af)_{,1}$ are defined for convenience.

For the metric (\ref{metric1}) we considered, the above Eqs. (\ref{eommatter31}-\ref{eommatter38}) reduce to two sets of equations
\begin{align}
&(e^{\psi+\mu_2}F_{12})_{,3}+(e^{\psi+\mu_3}F_{31})_{,2}=0,\nonumber\\
&(e^{\psi+\gamma}F_{01})_{,2}+(e^{\psi+\mu_2}F_{12})_{,0}=0,\nonumber\\
&(e^{\psi+\gamma}F_{01})_{,3}+(e^{\psi+\mu_3}F_{13})_{,0}=0,\label{oddmatter}\\
&(e^{\mu_2+\mu_3}P_{01})_{,0}+(e^{\gamma+\mu_3}P_{12})_{,2}+(e^{\gamma+\mu_2}P_{13})_{,3}\nonumber\\
&\;\;=e^{\psi+\mu_3}P_{02}Q_{02}+e^{\psi+\mu_2}P_{03}Q_{03}-e^{\psi+\gamma}P_{23}Q_{23}\nonumber
\end{align}
and
\begin{align}
&(e^{\psi+\mu_3}P_{02})_{,2}+(e^{\psi+\mu_2}P_{03})_{,3}=0,\nonumber\\
&(e^{\psi+\mu_2}P_{03})_{,0}-(e^{\psi+\gamma}P_{23})_{,2}=0,\nonumber\\
&(e^{\psi+\mu_3}P_{02})_{,0}+(e^{\psi+\gamma}P_{23})_{,3}=0,\label{evenmatter}\\
&(e^{\gamma+\mu_2}F_{02})_{,3}-(e^{\gamma+\mu_3}F_{03})_{,2}+(e^{\mu_2+\mu_3}F_{23})_{,0}\nonumber\\
&\;\;=e^{\psi+\gamma}F_{01}Q_{23}+e^{\psi+\mu_2}F_{12}Q_{03}-e^{\psi+\mu_3}F_{13}Q_{02}\nonumber.
\end{align}
Where the subscript $,i$ denotes taking partial derivative with respect to $x^i$. Note that the above equations are not independent, the first equation in each set is just the integrable condition of the following two equations. Note also that the set of equations (\ref{oddmatter}) involve only quantities which reverse their signs under the transformation $\varphi\rightarrow-\varphi$, while the set of equations (\ref{evenmatter}) involve only quantities that keep invariant under $\varphi\rightarrow-\varphi$. The quantities in (\ref{oddmatter}) and (\ref{evenmatter}) are called to be odd (or axial) and even (or polar) respectively.

\subsection{First order perturbation equations for metric and electromagnetic fields}
Now we consider perturbations of the field equations. In the above, we have noted that the EOMs of electromagnetic field can be splitted into two sets with just odd quantities  (\ref{oddmatter}) or just even quantities  (\ref{evenmatter}) respectively. The metric perturbations can also be divided into two sets, the perturbations $\sigma, q_2, q_3$ reverse their signs while the perturbations $\delta\gamma, \delta\psi, \delta\mu_2, \delta\mu_3$ keep invariant to keep the metric (\ref{metric1}) invariant under the transformation $\varphi\rightarrow-\varphi$, these two kinds of metric perturbations are also called to be odd and even respectively. From the metric (\ref{metric1}) it's obvious to see that the odd metric perturbations induce a dragging of inertial frame and impart a rotation of the black hole, while the even metric perturbations impart no such rotation.

In this paper, for the magnetic regular black holes we take the combination of the odd gravitational perturbations with even electromagnetic perturbations\cite{2004.07560}, i.e., we study the perturbed Einstein equations with odd parity caused by $\sigma, q_2, q_3$ combined with the perturbed electromagnetic equations with even parity (\ref{evenmatter}). This is different from the case discussed in Refs.\cite{2004.12185} and \cite{Chandrasekhar}, where for the electrically charged BHs, the authors considered odd gravitational perturbations combined with odd electromagnetic perturbations.

We perturb the Einstein equations
\begin{align}
R_{ab}&=-2\left(\mathcal{L}_FF_{am}F_b^{\;m}-\frac{1}{4}\eta_{ab}\mathcal{L}(F)\right),
\end{align}
to obtain
\begin{align}
\delta R_{ab}
=-4\mathcal{L}_{FF}\delta F_{pq}F^{pq}F_{am}F_b^{\;m}-2\mathcal{L}_F\delta F_{am}F_b^{\;m}-2\mathcal{L}_F F_{am}\delta F_b^{\;m}
+\eta_{ab}\mathcal{L}_F\delta F_{pq}F^{pq}.
\end{align}
To leading order of the perturbations, we have
\begin{align}
\delta R_{12}=-2\left(\mathcal{L}_FF_{13}\delta F_{23}\right),\;\;\;\;\;\delta R_{13}=0.\label{pertgrav}
\end{align}
The components of the strength tensor are given by
\begin{align}
F_{03}=\sigma e^{-\gamma}\frac{Q_m}{r}\sin\theta,\;\;\;\;\;\;F_{13}=\frac{Q_m}{r^2},\;\;\;\;\;\;F_{23}=q_2e^\gamma\frac{Q_m}{r}\sin\theta
\end{align}
with all other components vanishing. Note that the only non-vanishing component of the background strength tensor is $F_{13}$, the components $F_{03}$, $F_{23}$ and all other components are all small perturbations, which can be denoted as $\delta F_{03}$, $\delta F_{23}$, etc.

The linear perturbation version of the set of equations (\ref{evenmatter}) are given by
\begin{align}
&(e^{\psi+\mu_2}\delta P_{03})_{,0}-(e^{\psi+\gamma}\delta P_{23})_{,2}=0,\label{pertEM2}\\
&(e^{\psi+\mu_3}\delta P_{02})_{,0}+(e^{\psi+\gamma}\delta P_{23})_{,3}=0,\label{pertEM3}\\
&(e^{\gamma+\mu_2}\delta F_{02})_{,3}-(e^{\gamma+\mu_3}\delta F_{03})_{,2}+(e^{\mu_2+\mu_3}\delta F_{23})_{,0}=-e^{\psi+\mu_3}F_{13}Q_{02}.\label{pertEM4}
\end{align}
For the BH background (\ref{magbh}), taking derivative of  Eq.(\ref{pertEM4}) with respect to $x^0$ (or $t$) give rise to
\begin{align}
\left(\frac{\delta P_{02}}{\mathcal{L}_F}\right)_{,3,0}-\left(e^{\gamma}r \frac{\delta P_{03}}{\mathcal{L}_F}\right)_{,2,0}+\left(e^{-\gamma}r\frac{\delta P_{23}}{\mathcal{L}_F}\right)_{,0,0}=r^2\sin\theta F_{13}(q_{2,0,0}-\sigma_{,2,0}),\label{pertEM4n}
\end{align}
where we have replaced $\delta F_{ab}$ with $\delta P_{ab}$.
Substituting Eqs.(\ref{pertEM2}) and (\ref{pertEM3}) into Eq.(\ref{pertEM4n}), we have
\begin{align}
&\left[\frac{1}{\mathcal{L}_F}e^{-\psi-\mu_3}(e^{\psi+\gamma}\delta P_{23})_{,3}\right]_{,3}+\left[\frac{re^\gamma}{\mathcal{L}_F}e^{-\psi-\mu_2}(e^{\psi+\gamma}\delta P_{23})_{,2}\right]_{,2}\nonumber\\
&\;\;\;-\left(\frac{re^{-\gamma}}{\mathcal{L}_F}\delta P_{23}\right)_{,0,0}=-r^2\sin\theta F_{13}(q_{2,0,0}-\sigma_{,2,0}),
\end{align}
which, after inserting into the background metric (\ref{magbh}), can be rewritten as
\begin{align}
&\left[\frac{1}{\mathcal{L}_F}\frac{1}{r^2\sin\theta}(r\sin\theta e^{\gamma}\delta P_{23})_{,3}\right]_{,3}+\left[\frac{e^\gamma}{\mathcal{L}_F}(r e^{\gamma}\delta P_{23})_{,2}\right]_{,2}\nonumber\\
&\;\;\;-\left(\frac{re^{-\gamma}}{\mathcal{L}_F}\delta P_{23}\right)_{,0,0}=-r^2\sin\theta F_{13}(q_{2,0,0}-\sigma_{,2,0}).\label{pertEM4n3}
\end{align}

Considering the background metric (\ref{magbh}), the perturbed Einstein equations  (\ref{pertgrav}) can be written explicitly as
\begin{align}
&(r^2e^{2\gamma}Q_{23}\sin^3\theta)_{,3}+r^4Q_{02,0}\sin^3\theta
=2r^3e^\gamma\sin^2\theta\delta R_{12}
=4r^3e^\gamma\sin^2\theta \mathcal{L}_FF_{13}\delta F_{23},\label{pertgrav2}\\
&(r^2e^{2\gamma}Q_{23}\sin^3\theta)_{,2}-r^2e^{-2\gamma}Q_{03,0}\sin^3\theta
=-2r^2\sin^2\theta\delta R_{13}
=0.\label{pertgrav3}
\end{align}
Letting $Q(r,\theta,t)=r^2e^{2\gamma}Q_{23}\sin^3\theta$, Eqs.(\ref{pertgrav2}) and (\ref{pertgrav3}) can be rewritten as
\begin{align}
&\frac{\partial Q}{\partial\theta}-r^4(q_{2,0}-\sigma_{,2})_{,0}\sin^3\theta=4r^3e^\gamma\sin^2\theta F_{13}\delta P_{23},\label{pertgrav4}\\
&\frac{\partial Q}{\partial r}+r^2 e^{-2\gamma}(q_{3,0}-\sigma_{,3})_{,0}\sin^3\theta=0.\label{pertgrav5}
\end{align}
Eliminating $\sigma$ from the above two equations, we have
\begin{align}
\frac{\partial}{\partial r}\left(\frac{e^{2\gamma}}{r^2\sin^3\theta}\frac{\partial Q}{\partial r}\right)+\frac{\partial}{\partial \theta}\left(\frac{1}{r^4\sin^3\theta}\frac{\partial Q}{\partial \theta}\right)=Q_{23,0,0}
+\left(\frac{4e^\gamma}{r\sin\theta}F_{13}\delta P_{23}\right)_{,\theta}.
\end{align}
If the time dependence of the perturbation is assumed to be $Q(r,\theta,t)=\tilde{Q}(r, \theta)e^{i\omega t}$ and introduce $\Delta=r^2e^{2\gamma}$ and $B(r,\theta)=\delta P_{23}\sin\theta$ for convenience, the above equation can be rewritten as
\begin{align}
r^4\frac{\partial}{\partial r}\left(\frac{\Delta}{r^4}\frac{\partial \tilde{Q}}{\partial r}\right)+\sin^3\theta\frac{\partial}{\partial \theta}\left(\frac{1}{\sin^3\theta}\frac{\partial \tilde{Q}}{\partial \theta}\right)+\omega^2\frac{r^4}{\Delta}\tilde{Q}
=4r^3\sin^3\theta F_{13}e^\gamma \frac{\partial}{\partial \theta}\left(\frac{B}{\sin^2\theta}\right).\label{pertf21}
\end{align}
Combining Eqs.(\ref{pertEM4n3}) and (\ref{pertgrav4}) and eliminating  $(q_{2,0}-\sigma_{,2})_{,0}$, we have
\begin{align}
&\sin\theta\left[\frac{1}{\mathcal{L}_F}\frac{1}{r^2\sin\theta}(re^\gamma B)_{,\theta}\right]_{,\theta}
+\left[\frac{e^{2\gamma}}{\mathcal{L}_F}
(re^\gamma B)_{,r}\right]_{,r}\nonumber\\
&\;\;+\frac{1}{\mathcal{L}_F}r^2e^{-\gamma}\omega^2B=-\frac{1}{r^2\sin\theta}F_{13}\frac{\partial Q}{\partial\theta}+4re^\gamma B F_{13}^2.\label{pertf22}
\end{align}

Now, we separate the variables $r$ and $\theta$ in Eqs.(\ref{pertf21}) and (\ref{pertf22}). Suppose $\tilde{Q}(r,\theta)$ and $B(r,\theta)$ can be writen in the form
\begin{align}
&\tilde{Q}(r,\theta)=\bar{Q}(r)C_{l+2}^{-3/2}(\theta),\\
B(r,\theta)=&\frac{B(r)}{\sin\theta}\frac{d C_{l+2}^{-3/2}(\theta)}{d\theta}=3B(r)C_{l+1}^{-1/2}(\theta),\label{Brtheta}
\end{align}
where $C_{l+2}^{-3/2}(\theta)$, $C_{l+1}^{-1/2}(\theta)$ are Gegenbauer functions, and the relation
$\frac{1}{\sin\theta}\frac{dC_n^\rho}{d\theta}=-2\rho C_{n-1}^{\rho+1}$
has been used. Then, Eq.(\ref{pertf21}) becomes
\begin{align}
&r^4\frac{d}{d r}\left(\frac{\Delta}{r^4}\frac{d \bar{Q}}{d r}\right)C_{l+2}^{-3/2}(\theta)+\sin^3\theta\frac{d}{d \theta}\left(\frac{1}{\sin^3\theta}\frac{d C_{l+2}^{-3/2}(\theta)}{d \theta}\right)\bar{Q}(r)\nonumber\\
&+\omega^2\frac{r^4}{\Delta}\bar{Q}C_{l+2}^{-3/2}(\theta)
=4r^3\sin^3\theta F_{13}e^\gamma \frac{d}{d \theta}\left(\frac{1}{\sin^3\theta}\frac{d C_{l+2}^{-3/2}(\theta)}{d\theta}\right).\label{pertf211}
\end{align}
Applying the equation
\begin{align}
\left[\frac{d}{d\theta}\sin^{2\rho}\theta\frac{d}{d\theta}+n(n+2\rho)\sin^{2\rho}\theta\right]C_n^\rho(\theta)=0\label{Gegfunc}
\end{align}
satisfied by the Gegenbauer function, Eq.(\ref{pertf211}) reduces to
\begin{align}
&r^4\frac{d}{d r}\left(\frac{\Delta}{r^4}\frac{d \bar{Q}}{d r}\right)C_{l+2}^{-3/2}(\theta)-(l+2)(l-1)C_{l+2}^{-3/2}(\theta)\bar{Q}(r)\nonumber\\
&+\omega^2\frac{r^4}{\Delta}\bar{Q}C_{l+2}^{-3/2}(\theta)
=-4r^3\sin^3\theta F_{13}e^\gamma (l+2)(l-1) \frac{B(r)}{\sin^3 \theta}C_{l+2}^{-3/2}(\theta),\label{pertf212}
\end{align}
divided by $C_{l+2}^{-3/2}(\theta)$ on both sides of the Eq.(\ref{pertf212}), we have
\begin{align}
r^4\frac{d}{d r}\left(\frac{\Delta}{r^4}\frac{d \bar{Q}}{d r}\right)-(l+2)(l-1)\bar{Q}(r)
+\omega^2\frac{r^4}{\Delta}\bar{Q}
+4r^3 F_{13}e^\gamma (l+2)(l-1)B(r)=0.\label{pertf213}
\end{align}
For the Eq.(\ref{pertf22}), we substitute Eq.(\ref{Brtheta}) into it, which gives
\begin{align}
&\sin\theta\frac{B(r)}{r\mathcal{L}_F}\frac{d}{d\theta}\left[\frac{1}{\sin\theta}\frac{d}{d\theta}\left(\frac{1}{\sin\theta}\frac{d}{d\theta}
C_{l+2}^{-3/2}(\theta)\right)\right]+\frac{d}{dr}\left[\frac{e^{2\gamma}}{\mathcal{L}_F}\frac{d}{dr}\left(re^\gamma B(r)\right)\right]\nonumber\\
&\;+\frac{r^2}{\mathcal{L}_F}e^{-\gamma}\omega^2B(r)\frac{1}{\sin\theta}\frac{d C_{l+2}^{-3/2}}{d\theta}=-\frac{F_{13}}{r^2\sin\theta}
\frac{d C_{l+2}^{-3/2}}{d\theta}\bar{Q}(r)+4re^\gamma F_{13}^2B(r)\frac{1}{\sin\theta}\frac{d C_{l+2}^{-3/2}}{d\theta}.\label{pertf221}
\end{align}
Taking into account Eq.(\ref{Gegfunc}), the equation above can be further reduced to
\begin{align}
&\frac{1}{r\mathcal{L}_F}l(l+1)C_{l+1}^{-1/2}(\theta)B(r)+\frac{d}{dr}\left[\frac{e^{2\gamma}}{\mathcal{L}_F}\frac{d}{dr}\left(re^\gamma B(r)\right)
\right]C_{l+1}^{-1/2}(\theta)\nonumber\\
&\;+\omega^2r^2\frac{e^{-\gamma}}{\mathcal{L}_F}B(r)C_{l+1}^{-1/2}(\theta)=-\frac{F_{13}}{r^2}C_{l+1}^{-1/2}(\theta)\bar{Q}(r)+4re^\gamma F_{13}^2B(r)
C_{l+1}^{-1/2}(\theta).\label{pertf222}
\end{align}
Divided by $C_{l+1}^{-1/2}(\theta)$ on both sides of the Eq.(\ref{pertf222}), we have
\begin{align}
\frac{l(l+1)}{r\mathcal{L}_F} B(r)+\frac{d}{dr}\left[\frac{e^{2\gamma}}{\mathcal{L}_F}\frac{d}{dr}\left(re^\gamma B(r)\right)
\right]
+\omega^2r^2\frac{e^{-\gamma}}{\mathcal{L}_F}B(r)=-\frac{F_{13}}{r^2}\bar{Q}(r)+4re^\gamma F_{13}^2B(r).\label{pertf223}
\end{align}
Up to now, we have succeeded in separating variables of the master equations.

Next we need to write the master Eqs. (\ref{pertf213}) and (\ref{pertf223}) into the standard form. First we introduce the tortoise coordinate through
$dr_{*}=\frac{r^2}{\Delta}dr$,
and then we define two new functions $H_1(r)$ and $H_2(r)$
\begin{align}
\bar{Q}(r)=rH_1(r),\;\;\;\;\;\;\;r e^\gamma B(r)=\sqrt{\mathcal{L}_F}H_2(r),
\end{align}
these efforts enable us to rewrite  Eqs.(\ref{pertf213}) and (\ref{pertf223}) into the standard Sch\"odinger-type form finally
\begin{align}
&\left(\frac{d^2}{dr_{*}^2}+\omega^2\right)H_1+\left(\frac{\Delta \Delta'}{r^5}-\frac{4\Delta^2}{r^6}-(l+2)(l-1)\frac{\Delta}{r^4}\right)H_1\nonumber\\
&\;\;\;+4(l+2)(l-1)F_{13}\sqrt{\mathcal{L}_F}\frac{\Delta}{r^3}H_2=0,\label{meqG}\\
&\left(\frac{d^2}{dr_{*}^2}+\omega^2\right)H_2+\left(-l(l+1)\frac{\Delta}{r^4}-\frac{2\Delta^2}{r^5}\frac{(\sqrt{\mathcal{L}_F})'}
{\sqrt{\mathcal{L}_F}}-\frac{\Delta^2}{r^4}\frac{(\sqrt{\mathcal{L}_F})'\mathcal{L}_F'}{{\mathcal{L}_F}^{3/2}}\right.\nonumber\\
&\;\;\left.+\frac{\Delta'\Delta}{r^4}\frac{(\sqrt{\mathcal{L}_F})'}{\sqrt{\mathcal{L}_F}}+\frac{\Delta^2}{r^4}
\frac{(\sqrt{\mathcal{L}_F})''}{\sqrt{\mathcal{L}_F}}-\frac{4\Delta}{r^2}\mathcal{L}_FF_{13}^2\right)H_2+
\frac{\Delta}{r^3}F_{13}\sqrt{\mathcal{L}_F}H_1=0.\label{meqEM}
\end{align}

\section{Quasinormal modes\label{section3}}
In this section,  we will solve the perturbation equations (\ref{meqG}) and (\ref{meqEM}) to obtain QNMs $\omega$. There are five parameters $(\mu, \nu, M, q, \ell)$ in the equations. As an application and for simplicity, we fix $\mu=3$ and take $\nu=1, 2, 3$ which corresponds to the Maxwellian, Bardeen-like and Hayward-like solutions respectively. For these three types of solutions, the magnetic charge $q$ should lie in the range $q/M \in [0, 0.2963]$, $q/M \in [0, 0.7698]$ and $q/M \in [0, 1.0583]$ respectively \cite{Toshmatov:2019gxg}, for the existence of BH horizon. Also we fix the BH mass $M=1$ so other quantities are measured in units of $M$. For each value of these parameters, there exists two branches of frequencies, the electromagnetic branch and the gravitational branch, which we denote as $s_1$ and $s_2$ respectively. We will only consider $\ell \geq 2$ so both branches are present.

The two perturbation equations (\ref{meqG}) (\ref{meqEM}) are coupled and can be cast into a compact form as
\begin{align}
\left(\frac{d^2}{dr_{*}^2}+\omega^2-\mathbf{V}(r)\right) \mathbf{Y}=0, \label{MasterEq}
\end{align}
where $\mathbf{Y} = \left(\begin{array}{c}
    H_1 \\
    H_2
\end{array}\right)$ and $\mathbf{V}(r)$ is a $2\times 2$ matrix-valued function depending on $r$. To obtain QNMs, physical boundary conditions are necessary. Precisely, at the horizon and at infinity, the solution should take ingoing wave condition and outgoing wave condition respectively, that is
\begin{align}
& Y_i\sim b_i e^{-i\omega r_{*}} \qquad r\rightarrow r_{+},  \nonumber\\
& Y_i\sim  B_i e^{i\omega r_{*}} \qquad r\rightarrow \infty, \label{BoundaryConditions}
\end{align}
where $b_i, B_i$ are coefficients.

\begin{table}[!htbp]
    \begin{tabular}{|c|ll|ll|ll|}
    \hline
    \multirow{2}{*}{$q/M$} & \multicolumn{2}{c|}{$l=2$}     & \multicolumn{2}{c|}{$l=3$}     & \multicolumn{2}{c|}{$l=4$} \\ \cline{2-7}
     & \multicolumn{1}{c|}{$Ms_1$}      & \multicolumn{1}{c|}{$Ms_2$}     & \multicolumn{1}{c|}{$Ms_1$} & \multicolumn{1}{c|}{$Ms_2$}     & \multicolumn{1}{c|}{$Ms_1$}     & \multicolumn{1}{c|}{$Ms_2$}                                          \\ \hline
    0                  & \multicolumn{1}{l|}{\begin{tabular}[c]{@{}l@{}}0.457596 - \\0.0950045 $i$ \end{tabular}} & \begin{tabular}[c]{@{}l@{}}0.373672 - \\ 0.0889623 $i$ \end{tabular} & \multicolumn{1}{l|}{\begin{tabular}[c]{@{}l@{}} 0.656899 - \\ 0.0956164 $i$ \end{tabular}}  & \begin{tabular}[c]{@{}l@{}}0.599444 - \\ 0.0927032 $i$ \end{tabular}   & \multicolumn{1}{l|}{\begin{tabular}[c]{@{}l@{}}0.853095 -\\ 0.0958601 $i$\end{tabular}}  & \begin{tabular}[c]{@{}l@{}} 0.809178 -\\ 0.0941641 $i$ \end{tabular}  \\ \hline
    0.03                & \multicolumn{1}{l|}{\begin{tabular}[c]{@{}l@{}}0.482039 - \\0.0971335 $i$\end{tabular}}  & \begin{tabular}[c]{@{}l@{}}0.378366 - \\0.089435 $i$ \end{tabular}  & \multicolumn{1}{l|}{\begin{tabular}[c]{@{}l@{}} 0.68971 - \\0.096733 $i$\end{tabular}}  & \begin{tabular}[c]{@{}l@{}}0.607525 - \\0.0927656 $i$\end{tabular}  & \multicolumn{1}{l|}{\begin{tabular}[c]{@{}l@{}}0.895008 - \\0.09681 $i$\end{tabular}}  & \begin{tabular}[c]{@{}l@{}}0.822002 - \\0.0939324 $i$\end{tabular}  \\ \hline
    0.06                & \multicolumn{1}{l|}{\begin{tabular}[c]{@{}l@{}}0.50878 - \\0.098046 $i$\end{tabular}} & \begin{tabular}[c]{@{}l@{}}0.385304 - \\0.0904111 $i$\end{tabular}  & \multicolumn{1}{l|}{\begin{tabular}[c]{@{}l@{}}0.72499 - \\0.0991899 $i$\end{tabular}} & \begin{tabular}[c]{@{}l@{}}0.619409 - \\0.09402 $i$\end{tabular}  & \multicolumn{1}{l|}{\begin{tabular}[c]{@{}l@{}}0.936327 - \\0.0992792 $i$\end{tabular}}  & \begin{tabular}[c]{@{}l@{}}0.838917 - \\0.0948696 $i$\end{tabular}  \\ \hline
    0.10                & \multicolumn{1}{l|}{\begin{tabular}[c]{@{}l@{}}0.546865 - \\0.100171 $i$\end{tabular}}   & \begin{tabular}[c]{@{}l@{}}0.39726 - \\0.0916078 $i$\end{tabular}  & \multicolumn{1}{l|}{\begin{tabular}[c]{@{}l@{}}0.77456 - \\0.0999869 $i$\end{tabular}} & \begin{tabular}[c]{@{}l@{}}0.640635 - \\0.0951644 $i$\end{tabular}  & \multicolumn{1}{l|}{\begin{tabular}[c]{@{}l@{}}0.997253 - \\0.0998665 $i$\end{tabular}} & \begin{tabular}[c]{@{}l@{}}0.86938 - \\0.0968044 $i$\end{tabular}  \\ \hline
    0.13                & \multicolumn{1}{l|}{\begin{tabular}[c]{@{}l@{}}0.580935 - \\0.101569 $i$\end{tabular}}  & \begin{tabular}[c]{@{}l@{}}0.408263 - \\0.0919257 $i$\end{tabular} & \multicolumn{1}{l|}{\begin{tabular}[c]{@{}l@{}}0.818149 - \\0.101566 $i$\end{tabular}}  & \begin{tabular}[c]{@{}l@{}}0.659226 - \\0.0946591 $i$\end{tabular}  & \multicolumn{1}{l|}{\begin{tabular}[c]{@{}l@{}}1.04998 - \\0.101081 $i$\end{tabular}} & \begin{tabular}[c]{@{}l@{}}0.896203 - \\0.0959617 $i$\end{tabular}  \\ \hline
    0.16                & \multicolumn{1}{l|}{\begin{tabular}[c]{@{}l@{}}0.618284 - \\0.10222 $i$\end{tabular}}  & \begin{tabular}[c]{@{}l@{}}0.420734 - \\0.0916186 $i$\end{tabular} & \multicolumn{1}{l|}{\begin{tabular}[c]{@{}l@{}}0.865564 - \\0.102331 $i$\end{tabular}}  & \begin{tabular}[c]{@{}l@{}}0.679772 - \\0.094581 $i$\end{tabular} & \multicolumn{1}{l|}{\begin{tabular}[c]{@{}l@{}}1.1079 - \\0.102683 $i$\end{tabular}} & \begin{tabular}[c]{@{}l@{}}0.926126 - \\0.0969271 $i$\end{tabular} \\ \hline
    0.20               & \multicolumn{1}{l|}{\begin{tabular}[c]{@{}l@{}}0.679892 - \\0.102856 $i$\end{tabular}} & \begin{tabular}[c]{@{}l@{}}0.440005 - \\0.0909658 $i$\end{tabular} & \multicolumn{1}{l|}{\begin{tabular}[c]{@{}l@{}}0.943037 - \\0.102667 $i$\end{tabular}}  & \begin{tabular}[c]{@{}l@{}}0.715085 - \\0.0958283 $i$\end{tabular}  & \multicolumn{1}{l|}{\begin{tabular}[c]{@{}l@{}}1.20169 - \\0.102647 $i$\end{tabular}}  & \begin{tabular}[c]{@{}l@{}}0.977831 - \\0.0962744 $i$\end{tabular}  \\ \hline
    \end{tabular}
    \caption{Fundamental modes for $\nu=1$.} \label{nu1}
    \end{table}

    \begin{table}[!htbp]
    \begin{tabular}{|c|ll|ll|ll|}
    \hline
    \multirow{2}{*}{$q/M$} & \multicolumn{2}{c|}{$l=2$}     & \multicolumn{2}{c|}{$l=3$}     & \multicolumn{2}{c|}{$l=4$} \\ \cline{2-7}
     & \multicolumn{1}{c|}{$Ms_1$}      & \multicolumn{1}{c|}{$Ms_2$}     & \multicolumn{1}{c|}{$Ms_1$} & \multicolumn{1}{c|}{$Ms_2$}     & \multicolumn{1}{c|}{$Ms_1$}     & \multicolumn{1}{c|}{$Ms_2$}                                          \\ \hline
    0                  & \multicolumn{1}{l|}{\begin{tabular}[c]{@{}l@{}}0.457596 - \\0.0950045 $i$ \end{tabular}} & \begin{tabular}[c]{@{}l@{}}0.373672 - \\ 0.0889623 $i$ \end{tabular} & \multicolumn{1}{l|}{\begin{tabular}[c]{@{}l@{}} 0.656899 - \\ 0.0956164 $i$ \end{tabular}}  & \begin{tabular}[c]{@{}l@{}}0.599444 - \\ 0.0927032 $i$ \end{tabular}   & \multicolumn{1}{l|}{\begin{tabular}[c]{@{}l@{}}0.853095 -\\  0.0958601 $i$\end{tabular}}  & \begin{tabular}[c]{@{}l@{}} 0.809178 -\\ 0.0941641 $i$ \end{tabular}  \\ \hline
    0.1                & \multicolumn{1}{l|}{\begin{tabular}[c]{@{}l@{}}0.468975 - \\0.0966194 $i$\end{tabular}}  & \begin{tabular}[c]{@{}l@{}}0.373045 - \\0.0885194 $i$ \end{tabular}  & \multicolumn{1}{l|}{\begin{tabular}[c]{@{}l@{}} 0.666858 - \\0.0957702 $i$\end{tabular}}  & \begin{tabular}[c]{@{}l@{}}0.599029 - \\0.0918889 $i$\end{tabular}  & \multicolumn{1}{l|}{\begin{tabular}[c]{@{}l@{}}0.861766 - \\0.0970081 $i$\end{tabular}}  & \begin{tabular}[c]{@{}l@{}}0.809553 - \\0.0937172 $i$\end{tabular}  \\ \hline
    0.2                & \multicolumn{1}{l|}{\begin{tabular}[c]{@{}l@{}}0.474839 - \\0.0965579 $i$\end{tabular}} & \begin{tabular}[c]{@{}l@{}}0.37312 - \\0.0882093 $i$\end{tabular}  & \multicolumn{1}{l|}{\begin{tabular}[c]{@{}l@{}}0.67422 - \\0.0954732 $i$\end{tabular}} & \begin{tabular}[c]{@{}l@{}}0.598863 - \\0.0915312 $i$\end{tabular}  & \multicolumn{1}{l|}{\begin{tabular}[c]{@{}l@{}}0.871742 - \\0.0967447 $i$\end{tabular}}  & \begin{tabular}[c]{@{}l@{}}0.809297 - \\0.0932967 $i$\end{tabular}  \\ \hline
    0.3                & \multicolumn{1}{l|}{\begin{tabular}[c]{@{}l@{}}0.48484 - \\0.0962131 $i$\end{tabular}}   & \begin{tabular}[c]{@{}l@{}}0.373449 - \\0.0876668 $i$\end{tabular}  & \multicolumn{1}{l|}{\begin{tabular}[c]{@{}l@{}}0.686297 - \\0.095169 $i$\end{tabular}} & \begin{tabular}[c]{@{}l@{}}0.599205 - \\0.0909092 $i$\end{tabular}  & \multicolumn{1}{l|}{\begin{tabular}[c]{@{}l@{}}0.887226 - \\0.0954353 $i$\end{tabular}} & \begin{tabular}[c]{@{}l@{}}0.810104 - \\0.0925481 $i$\end{tabular}  \\ \hline
    0.4                & \multicolumn{1}{l|}{\begin{tabular}[c]{@{}l@{}}0.499273 - \\0.0953048 $i$\end{tabular}}  & \begin{tabular}[c]{@{}l@{}}0.374301 - \\0.0868343 $i$\end{tabular} & \multicolumn{1}{l|}{\begin{tabular}[c]{@{}l@{}}0.703836 - \\0.0952043 $i$\end{tabular}}  & \begin{tabular}[c]{@{}l@{}}0.600697 -\\ 0.0899732 $i$\end{tabular}  & \multicolumn{1}{l|}{\begin{tabular}[c]{@{}l@{}}0.907249 -\\ 0.09402 $i$\end{tabular}} & \begin{tabular}[c]{@{}l@{}}0.812951 -\\ 0.0914434 $i$\end{tabular}  \\ \hline
    0.5                & \multicolumn{1}{l|}{\begin{tabular}[c]{@{}l@{}}0.518875 -\\ 0.0937642 $i$\end{tabular}}  & \begin{tabular}[c]{@{}l@{}}0.376017 -\\ 0.085577 $i$\end{tabular} & \multicolumn{1}{l|}{\begin{tabular}[c]{@{}l@{}}0.729105 -\\ 0.0943042 $i$\end{tabular}}  & \begin{tabular}[c]{@{}l@{}}0.604038 -\\ 0.0885838 $i$\end{tabular} & \multicolumn{1}{l|}{\begin{tabular}[c]{@{}l@{}}0.935047 - \\0.0943714 $i$\end{tabular}} & \begin{tabular}[c]{@{}l@{}}0.818769 -\\ 0.0898618 $i$\end{tabular} \\ \hline
    0.6               & \multicolumn{1}{l|}{\begin{tabular}[c]{@{}l@{}}0.546067 -\\0.0914386 $i$\end{tabular}} & \begin{tabular}[c]{@{}l@{}}0.37901 -\\0.0835757 $i$\end{tabular} & \multicolumn{1}{l|}{\begin{tabular}[c]{@{}l@{}}0.761723 -\\ 0.0906819 $i$\end{tabular}}  & \begin{tabular}[c]{@{}l@{}}0.610073 -\\ 0.0863319 $i$\end{tabular}  & \multicolumn{1}{l|}{\begin{tabular}[c]{@{}l@{}}0.974356 -\\ 0.0903898 $i$\end{tabular}}  & \begin{tabular}[c]{@{}l@{}}0.828703 -\\ 0.087355 $i$\end{tabular}  \\ \hline
    \end{tabular}
    \caption{Fundamental modes for $\nu=2$.} \label{nu2}
    \end{table}

    \begin{table}[!htbp]
    \begin{tabular}{|c|ll|ll|ll|}
    \hline
    \multirow{2}{*}{$q/M$} & \multicolumn{2}{c|}{$l=2$}                                                                                                                                   & \multicolumn{2}{c|}{$l=3$}                                                                                                                                   & \multicolumn{2}{c|}{$l=4$}                                                                                                                                   \\ \cline{2-7}
                       & \multicolumn{1}{c|}{$Ms_1$}                                                               & \multicolumn{1}{c|}{$Ms_2$}                                          & \multicolumn{1}{c|}{$Ms_1$}                                                               & \multicolumn{1}{c|}{$Ms_2$}                                          & \multicolumn{1}{c|}{$Ms_1$}                                                               & \multicolumn{1}{c|}{$Ms_2$}                                          \\ \hline
    0                  & \multicolumn{1}{l|}{\begin{tabular}[c]{@{}l@{}}0.457596 - \\ 0.0950045 $i$\end{tabular}} & \begin{tabular}[c]{@{}l@{}}0.373672 - \\ 0.0889623 $i$\end{tabular} & \multicolumn{1}{l|}{\begin{tabular}[c]{@{}l@{}}0.656899 -\\ 0.0956164 $i$\end{tabular}}  & \begin{tabular}[c]{@{}l@{}}0.599444 -\\ 0.0927032 $i$\end{tabular}   & \multicolumn{1}{l|}{\begin{tabular}[c]{@{}l@{}}0.853095 -\\ 0.0958601 $i$\end{tabular}}  & \begin{tabular}[c]{@{}l@{}}0.809178 -\\ 0.0941641 $i$\end{tabular}  \\ \hline
    0.2                & \multicolumn{1}{l|}{\begin{tabular}[c]{@{}l@{}}0.484066 -\\ 0.0971561 $i$\end{tabular}}  & \begin{tabular}[c]{@{}l@{}}0.373029 -\\ 0.0885771 $i$\end{tabular}  & \multicolumn{1}{l|}{\begin{tabular}[c]{@{}l@{}}0.676035 -\\ 0.0958228 $i$\end{tabular}}  & \begin{tabular}[c]{@{}l@{}}0.599096 -\\ 0.0919502 $i$\end{tabular}  & \multicolumn{1}{l|}{\begin{tabular}[c]{@{}l@{}}0.868435 -\\ 0.0969739 $i$\end{tabular}}  & \begin{tabular}[c]{@{}l@{}}0.809700 -\\ 0.0937901 $i$\end{tabular}  \\ \hline
    0.4                & \multicolumn{1}{l|}{\begin{tabular}[c]{@{}l@{}}0.487026 - \\ 0.0967646 $i$\end{tabular}} & \begin{tabular}[c]{@{}l@{}}0.372949 -\\ 0.0882681 $i$\end{tabular}  & \multicolumn{1}{l|}{\begin{tabular}[c]{@{}l@{}}0.679654 - \\ 0.0954973 $i$\end{tabular}} & \begin{tabular}[c]{@{}l@{}}0.598688 -\\ 0.0915619 $i$\end{tabular}  & \multicolumn{1}{l|}{\begin{tabular}[c]{@{}l@{}}0.873348 -\\ 0.0964836 $i$\end{tabular}}  & \begin{tabular}[c]{@{}l@{}}0.808999 -\\ 0.0933831 $i$\end{tabular}  \\ \hline
    0.6                & \multicolumn{1}{l|}{\begin{tabular}[c]{@{}l@{}}0.495228 -\\ 0.095471 $i$\end{tabular}}   & \begin{tabular}[c]{@{}l@{}}0.372795 -\\ 0.0873902 $i$\end{tabular}  & \multicolumn{1}{l|}{\begin{tabular}[c]{@{}l@{}}0.689621 - \\ 0.0946848 $i$\end{tabular}} & \begin{tabular}[c]{@{}l@{}}0.597882 -\\ 0.0904792 $i$\end{tabular}  & \multicolumn{1}{l|}{\begin{tabular}[c]{@{}l@{}}0.885956 - \\ 0.0948336 $i$\end{tabular}} & \begin{tabular}[c]{@{}l@{}}0.807840 -\\ 0.0922095 $i$\end{tabular}  \\ \hline
    0.8                & \multicolumn{1}{l|}{\begin{tabular}[c]{@{}l@{}}0.512083 -\\ 0.0921845 $i$\end{tabular}}  & \begin{tabular}[c]{@{}l@{}}0.372704 - \\ 0.0854734 $i$\end{tabular} & \multicolumn{1}{l|}{\begin{tabular}[c]{@{}l@{}}0.710699 -\\ 0.0926479 $i$\end{tabular}}  & \begin{tabular}[c]{@{}l@{}}0.597269 -\\ 0.0881798 $i$\end{tabular}  & \multicolumn{1}{l|}{\begin{tabular}[c]{@{}l@{}}0.909488 - \\ 0.0917749 $i$\end{tabular}} & \begin{tabular}[c]{@{}l@{}}0.807622 -\\ 0.0896405 $i$\end{tabular}  \\ \hline
    0.9                & \multicolumn{1}{l|}{\begin{tabular}[c]{@{}l@{}}0.525768 -\\ 0.0891379 $i$\end{tabular}}  & \begin{tabular}[c]{@{}l@{}}0.372739 - \\ 0.0838454 $i$\end{tabular} & \multicolumn{1}{l|}{\begin{tabular}[c]{@{}l@{}}0.727753 -\\ 0.0896736 $i$\end{tabular}}  & \begin{tabular}[c]{@{}l@{}}0.597416 - \\ 0.0862594 $i$\end{tabular} & \multicolumn{1}{l|}{\begin{tabular}[c]{@{}l@{}}0.928251 - \\ 0.0896580 $i$\end{tabular}} & \begin{tabular}[c]{@{}l@{}}0.808613 - \\ 0.0874729 $i$\end{tabular} \\ \hline
    0.95               & \multicolumn{1}{l|}{\begin{tabular}[c]{@{}l@{}}0.534733 - \\ 0.0868849 $i$\end{tabular}} & \begin{tabular}[c]{@{}l@{}}0.372759 - \\ 0.0827797 $i$\end{tabular} & \multicolumn{1}{l|}{\begin{tabular}[c]{@{}l@{}}0.738341 -\\ 0.0872509 $i$\end{tabular}}  & \begin{tabular}[c]{@{}l@{}}0.597685 -\\ 0.0849986 $i$\end{tabular}  & \multicolumn{1}{l|}{\begin{tabular}[c]{@{}l@{}}0.940533 -\\ 0.0877607 $i$\end{tabular}}  & \begin{tabular}[c]{@{}l@{}}0.809554 -\\ 0.0860444 $i$\end{tabular}  \\ \hline
    \end{tabular}
    \caption{Fundamental modes for $\nu=3$.} \label{nu3}
    \end{table}

\begin{figure}[!htbp]
\includegraphics[width=.32\textwidth]{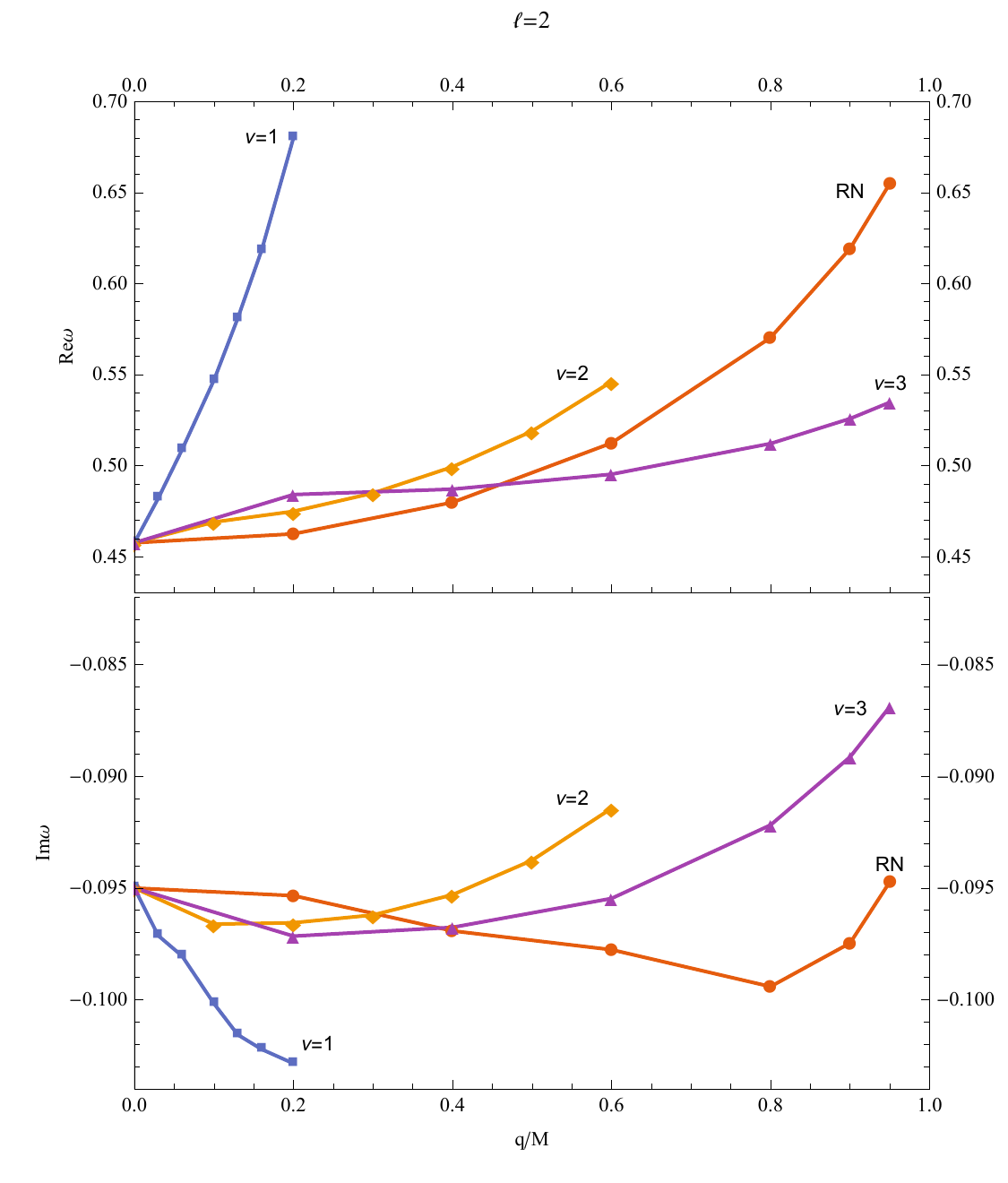}
\includegraphics[width=.32\textwidth]{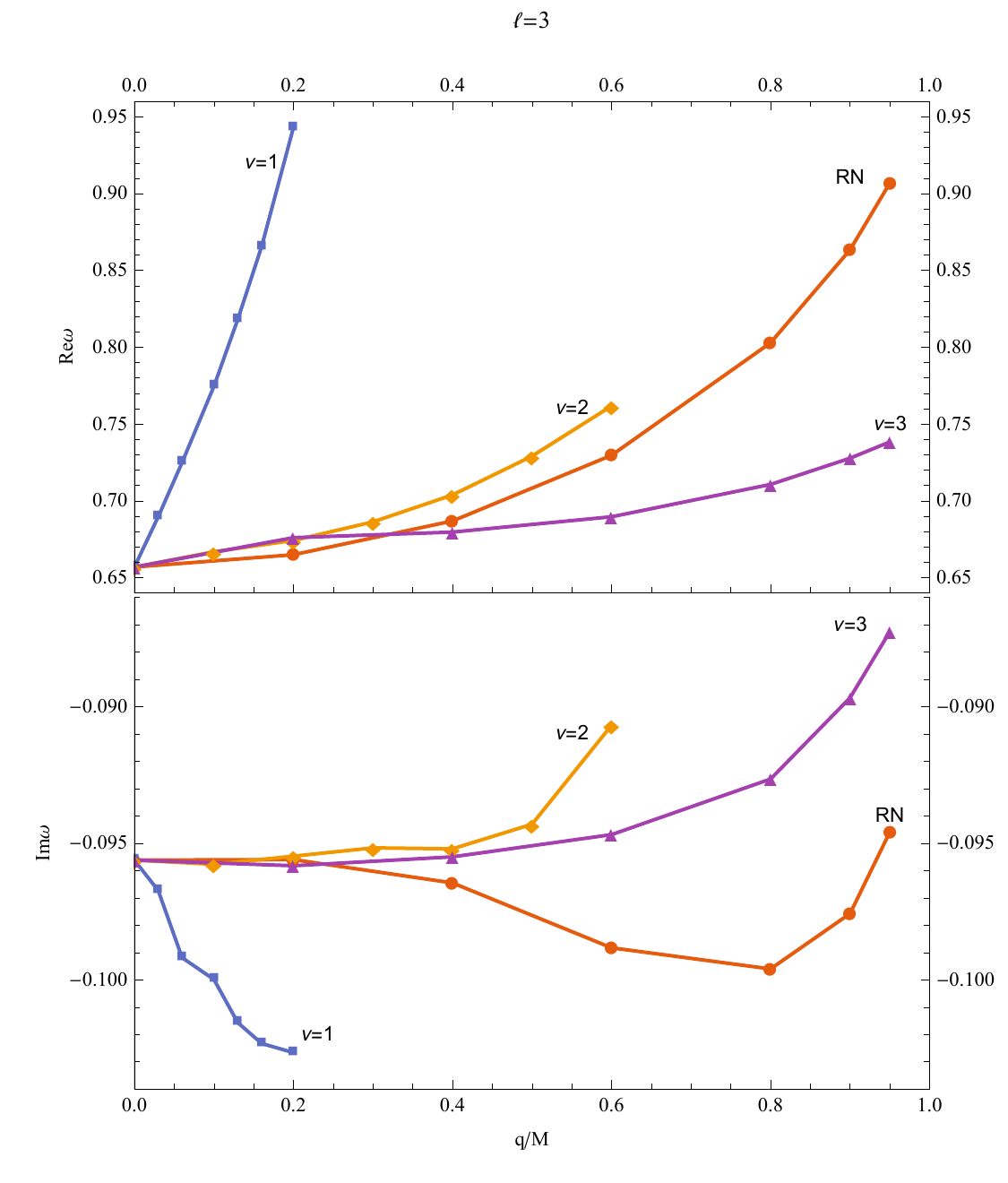}
\includegraphics[width=.32\textwidth]{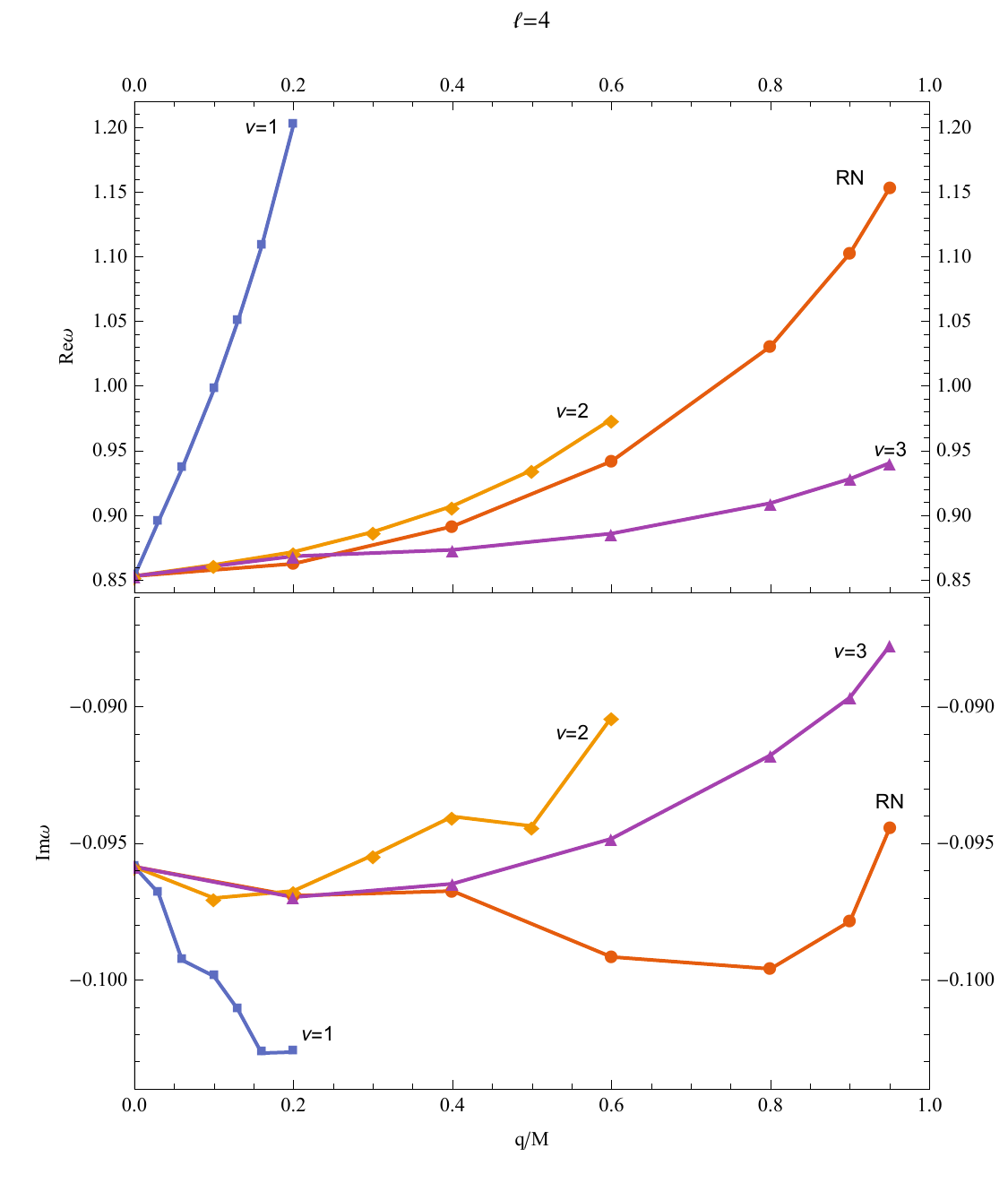}
\caption{Fundamental modes of electromagnetic branch $M s_1$.}
\label{ElectromagneticQNMs}
\end{figure}

\begin{figure}[!htbp]
\includegraphics[width=0.32\textwidth]{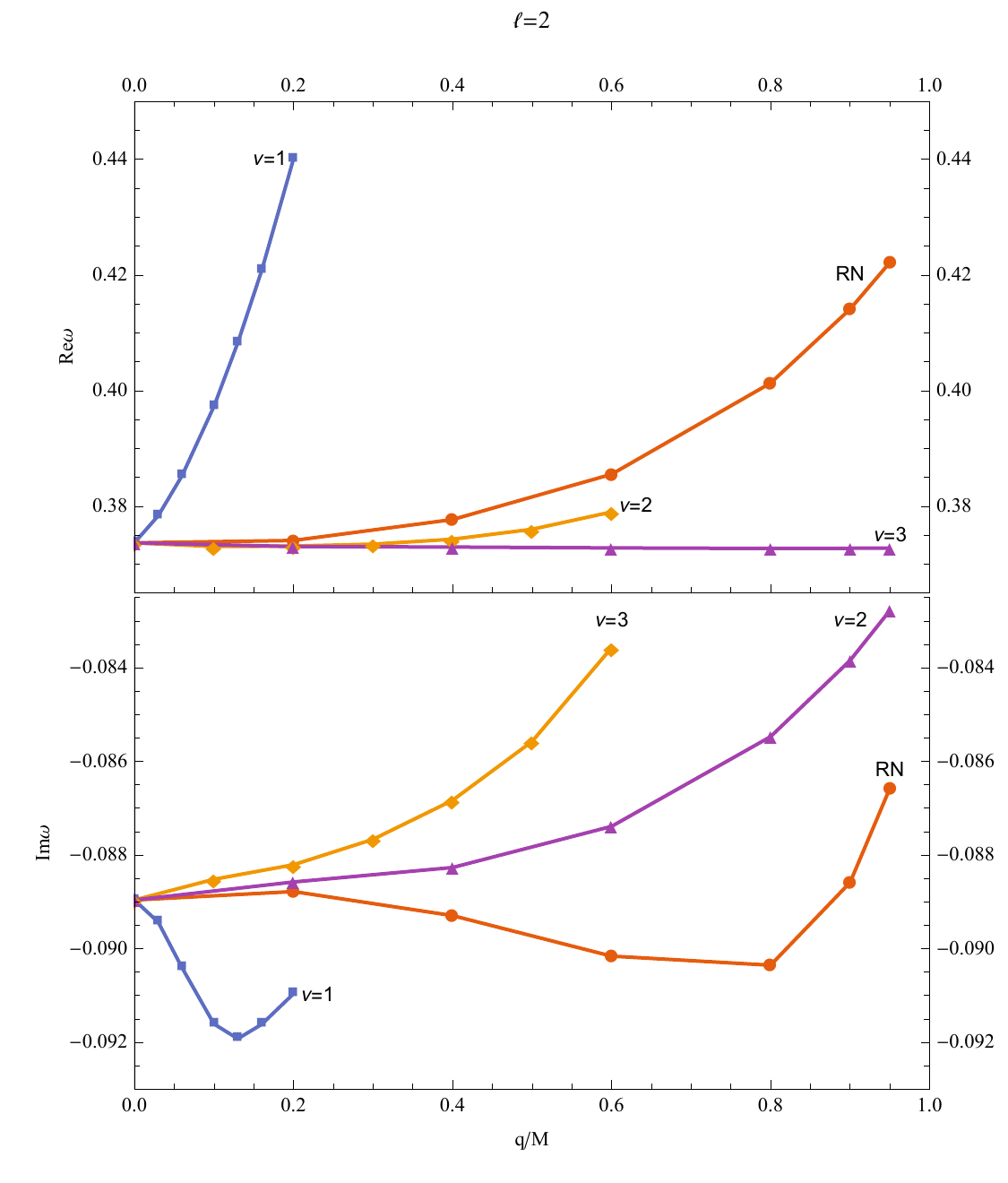}
\includegraphics[width=0.32\textwidth]{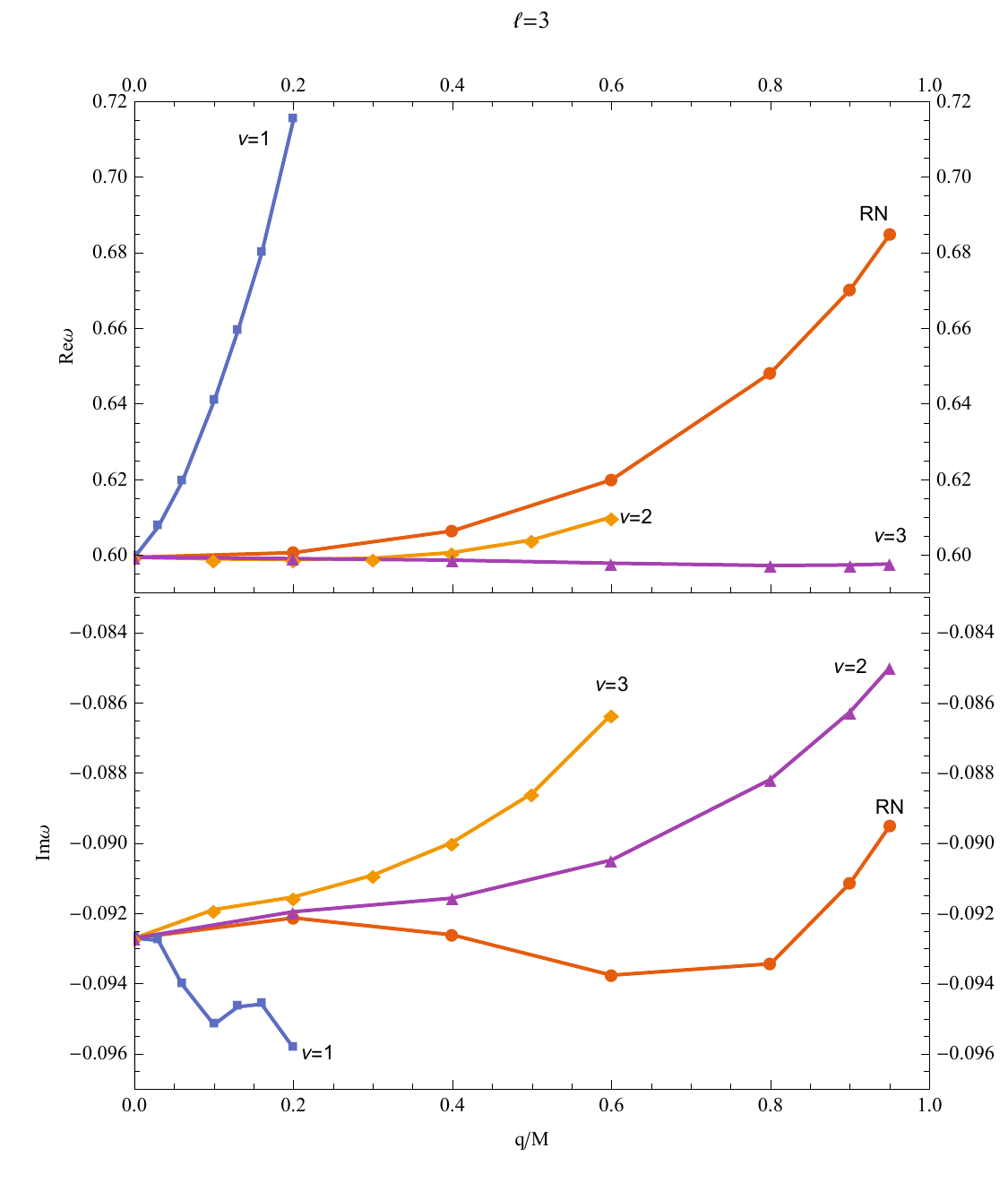}
\includegraphics[width=0.32\textwidth]{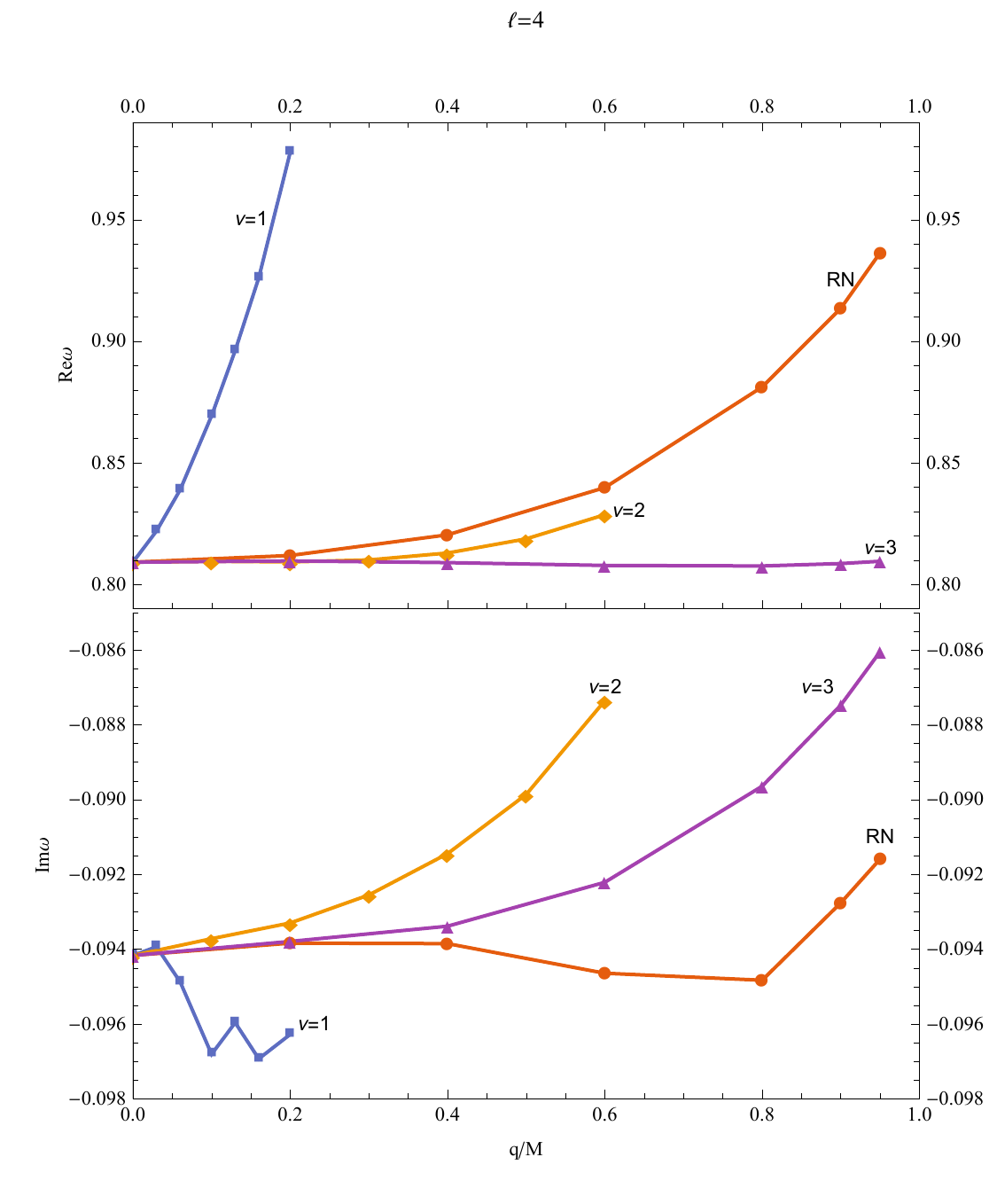}
\caption{Fundamental modes of gravitational branch $M s_2$.}
\label{GravitaionalQNMs}
\end{figure}

With the above boundary conditions, it is now an eigenvalue problem to obtain QNMs $\omega$ from the perturbation equation (\ref{MasterEq}), which can be solved by applying the matrix-valued direct integration method. For details on this numerical method, readers can refer to ref. \cite{Pani:2013pma}. Numerical results of fundamental modes, which are expected to dominate the late time of ringdown stage, are shown in Table \ref{nu1}, \ref{nu2} and \ref{nu3} for various values of $q$. For more visual comparison, we also plot the numerical data in Fig.\ref{GravitaionalQNMs} and \ref{ElectromagneticQNMs}. Gravito-electromagnetic QNMs of RN BH carrying  magnetic charge in GR are also shown for comparison. From the tables and figures, following silent properties of QNMs can be observed:
\begin{itemize}
    \item When $q = 0$, the BH solution (\ref{magbh}) reduces to the well-known Schwarzschild BH. And the perturbation equations (\ref{meqG}) and (\ref{meqEM}) reduces to that of gravito-electromagnetic perturbations of the Schwarzschild BH in GR. So in this case, QNMs of the three types of regular BHs and RN BH coincide, which confirms the validity of our numerical method.

    \item When $q \neq 0$, QNMs of the three types of regular BHs depend significantly on the parameter $\nu$ of the theory and the magnetic charge $q$, and are very different from that of the RN BH. This may give us a way to distinguish different theories and BHs.

    \item In the parameter space we considered, all the QNMs have negative imaginary part ${\rm Im} \omega < 0$ suggesting that these BHs are stable under gravito-electromagnetic perturbations.

    \item For the electromagnetic branch $M s_1$, with the increase of the magnetic charge $q$, the real part of QNMs ${\rm Re} \omega$ increases monotonically for all the three types of regular BHs. Meanwhile, the dependence of the imaginary part of QNMs ${\rm Im} \omega$ on $q$ is complicated relying on the value of $\ell$. For $\ell=2$ mode which is believed to dominate the ringdown stage,  with the increase of $q$, ${\rm Im} \omega$ of $\nu=1$ type decreases monotonically, while that of $\nu=2$ or $\nu=3$ type first decreases then increases. This means that $\nu=1$ type BH becomes more stable when $q$ becomes larger.

    \item For the gravitational branch $M s_2$, with the increase of the charge $q$, ${\rm Re} \omega$ of either $\nu=1$ or $\nu=2$ type increases monotonically, while that of $\nu=3$ type is nearly unaffected. Meanwhile, ${\rm Im} \omega$ of either $\nu=2$ or $\nu=3$ increases monotonically, while that of $\nu=1$ depends on $q$ in a complicated way. This means that $\nu=2$ and $\nu=3$ type BHs becomes less stable when $q$ becomes larger.

    \item It is more interesting to see the effect of the value of $\nu$ on ${\rm Im} \omega$. From the figures, one can see that in both branches, for fixed $q$, QNMs of $\nu=1$ type always have the most negative ${\rm Im} \omega$ meaning that the Maxwellian BH is the most stable among the three types of regular BHs and also more stable than RN BH.
\end{itemize}

\section{Summary\label{section4}}

In this work, we studied the gravito-electromagnetic perturbations of magnetic regular BHs. By applying Chandrasekha's approach, the master equations are obtained in which gravitational perturbations with odd-parity are coupled with the electromagnetic perturbations with even-parity. These master equations can be used to derive the QNMs of such kind of magnetic regular BHs. As an application, we fixed $\mu=3$ and take $\nu=1, 2$ and $3$ which correspond to the new Maxwellian, Bardeen-like and Hayward-like BHs respectively constructed in \cite{1610.02636}, and calculate the corresponding QNMs by applying numerical method. From the results, one can see that the QNMs depends significantly on the value of $\nu$ and the magnetic charge $q$ and are very different from that of RN BH carrying magnetic charge. In the parameter space we considered, all the QNMs have negative imaginary part ${\rm Im} \omega$ suggesting that these BHs are all stable under the perturbations. By comparing the three types of regular BHs and the standard magnetic RN BH, it is interesting to see that QNMs of $\nu=1$ type always have the most negative ${\rm Im} \omega$ indicating that the Maxwellian BH is most stable. These properties of QNMs give us a way to distinguish the different theories and BHs.

Of course, there remains several questions for further investigations. First, in this work we only consider the gravitational perturbations with odd-parity and the electromagnetic perturbations with even-parity. It is interesting to include the other half of perturbations, namely the gravitational perturbations with even-parity and the electromagnetic perturbations with odd-parity, which will give us a more complete picture of the perturbations and more information about the QNMs. Second, in applications of the master equations in this work, we only consider the case with $\mu=3$ for simplicity. It is straightforward to extend our discussions to other values of $\mu$ to see its effects on the QNMs. Last but not least, it will also be interesting to extend the method to discuss perturbations of other kind of regular BHs in other theories.

\section*{Acknowledgment}

KM would like to thank Profs. R.Konoplya and O.Sarbach for valuable discussions. SJZ is supported by the National Natural Science Foundation of China (NNSFC) under Grant No. 12075207.

\providecommand{\href}[2]{#2}\begingroup
\footnotesize\itemsep=0pt
\providecommand{\eprint}[2][]{\href{http://arxiv.org/abs/#2}{arXiv:#2}}

\end{document}